# AVERAGE OPTIMALITY FOR RISK-SENSITIVE CONTROL WITH GENERAL STATE SPACE[1]

By Anna Jaśkiewicz

*Wrocław University of Technology*

This paper deals with discrete-time Markov control processes on a general state space. A long-run risk-sensitive average cost criterion is used as a performance measure. The one-step cost function is nonnegative and possibly unbounded. Using the vanishing discount factor approach, the optimality inequality and an optimal stationary strategy for the decision maker are established.

**1. Introduction and the model.** This paper deals with discrete-time Markov control processes on a general state space. The one-step cost function is nonnegative and possibly unbounded. The decision maker is supposed to be risk-averse with a constant risk coefficient $\gamma > 0$. The risk-sensitive average cost criterion is used as a performance measure. The aim of the work is to establish the optimality inequality for risk-sensitive dynamic programming and derive an optimal stationary policy. The result is proved under two different sets of compactness-continuity assumptions, namely, for Markov control processes with weakly continuous transition probabilities [Condition (**W**)], as well as transition probabilities that are continuous with respect to setwise convergence [Condition (**S**)]. A similar problem for risk-neutral stochastic control models has been examined in [27] using the vanishing discount factor approach. However, it is well known that, for risk-sensitive control models, an analogous approximation of the average cost via a sequence of the corresponding discounted models does not work. Instead of this, following [9, 15, 16], we introduce an auxiliary discounted minimax problem. A variational formula that expresses the mutual relationship between the relative entropy function and the logarithmic moment-generating function enables us to connect the discounted minimax model with the original one.

Received March 2006; revised September 2006.
[1]Supported by MEiN Grant 1 P03A 01030.

Next, assuming that a certain family of functions is bounded [Condition (**B**)] and using Fatou's lemma (for weakly or setwise convergent measures), we obtain the optimality inequality.

The predecessor of our result is Theorem 4.1 in [16], where the optimality inequality for the risk-sensitive dynamic programming with a countable state space was established. Instead of boundedness assumption (**B**), Hernández-Hernández and Marcus [16] assume that there exists a stationary policy which induces a finite average cost that is equal some constant in each state. On the other hand, it is well known that an optimal risk-sensitive average cost may depend on the initial state (see Example 1). This behavior happens if the risk factor is too large. Instead of this restriction on the risk coefficient, we use Condition (**B**), which makes the process reach "good states" sufficiently fast.

There is a rich literature in risk-sensitive control, going back at least to the seminal works of Howard and Matheson [18] and Jacobson [19], which covered the finite horizon case. The average cost criterion on the infinite horizon was studied in [5, 8, 14, 15, 16, 31] for a denumerable state space and in [10, 11, 20] for a general state space. It is also worth mentioning that risk-sensitive control finds natural applications in portfolio managment, where the objective is to maximize the growth rate of the expected utility of wealth; see [3, 4, 30] and the references cited therein.

The paper is organized as follows. Below a Markov control model with the long-run average cost criterion as a performance measure is described, as well as some basic notation is set up. In Section 2 we introduce preliminaries and present the auxiliary discounted minimax problem, which is, in turn, solved in Section 3. The main result is established in Section 4. Section 5 contains a discussion of Condition (**B**), and in the Appendix a variational formula for the logarithmic moment-generating function is stated.

A discrete-time Markov control process is specified by the following objects:

(i) The *state space* $X$ is a standard Borel space (i.e., a nonempty Borel subset of some Polish space).

(ii) $A$ is a Borel *action space*.

(iii) $K$ is a nonempty Borel subset of $X \times A$. We assume that, for each $x \in X$, the nonempty $x$-section

$$A(x) = \{a \in A : (x,a) \in K\}$$

of $K$ is compact and represents the *set of actions available* in state $x$.

(iv) $q$ is a regular *conditional distribution* from $K$ to $X$.

(v) The one-step *cost function* $c$ is a Borel measurable mapping from $K$ to $[0, +\infty]$.

Then the history spaces are defined as $H_0 = X$, $H_k = (X \times A)^k \times X$ and $H_\infty = (X \times A)^\infty$. As usual, a *policy* $\pi = \{\pi_k, k = 0, 1, \ldots\} \in \Pi$ is a sequence of transition probabilities from $H_k$ to $A$ such that $\pi_k(A(x_k)|h_k) = 1$, where $h_k = (x_0, a_0, \ldots, x_k) \in H_k$. The class of *stationary policies* is identified with the class $F$ of measurable functions $f$ from $X$ to $A$ such that $f(x) \in A(x)$. It is well known that $F$ is nonempty [6]. By the Ionescu–Tulcea theorem [24], for each policy $\pi$ and each initial state $x_0 = x$, a probability measure $\mathbf{P}_x^\pi$ and a stochastic process $\{(x_k, a_k)\}$ are defined on $H_\infty$ in a canonical way, where $x_k$ and $a_k$ describe the state and the decision at stage $k$, respectively. By $E_x^\pi$ we denote the expectation operator with respect to the probability measure $\mathbf{P}_x^\pi$.

Let $\gamma > 0$ be a given risk factor. For any initial state $x \in X$ and policy $\pi \in \Pi$, we define the following risk-sensitive average cost criterion:

$$J(x, \pi) = \limsup_{n \to \infty} \frac{1}{\gamma n} \log E_x^\pi \exp\left\{\gamma \sum_{k=0}^{n-1} c(x_k, a_k)\right\}.$$

Our aim is to minimize $J(x, \pi)$ within the class of all policies and find a policy $\pi^*$, for which

$$J^*(x) := \inf_{\pi \in \Pi} J(x, \pi) = J(x, \pi^*).$$

Throughout the paper the following assumption will be supposed to hold true even without explicit reference:

(G) $\qquad \exists \widetilde{\pi} \in \Pi \quad J(x, \widetilde{\pi}) < +\infty.$

REMARK 1. Throughout the remainder, we assume that the risk factor $\gamma > 0$ is *arbitrary* and *fixed*. Therefore, here and subsequently, we shall not indicate that some quantities depend on $\gamma$ [e.g., we write $J(x, \pi)$ instead of $J^\gamma(x, \pi)$, dropping the index $\gamma$].

**2. Preliminaries.** Let $\Pr(X)$ be the set of all probability measures on $X$. Fix $\nu \in \Pr(X)$. The relative entropy function $R(\cdot \| \nu)$ is a mapping from $\Pr(X)$ into $\mathbb{R}$ defined as follows:

$$R(\mu \| \nu) := \begin{cases} \int_X \log \frac{d\mu}{d\nu} d\mu, & \mu \ll \nu, \\ +\infty, & \text{otherwise.} \end{cases}$$

It is well known that $R(\mu \| \nu)$ is nonnegative for any $\mu \in \Pr(X)$ and $R(\mu \| \nu) = 0$ if and only if $\mu = \nu$ (consult Lemma 1.4.1 in [12]).

We shall consider the following auxiliary minimax problem, associated with our original Markov control process. The set $X$ is the state space, while $A$ and $\Pr(X)$ are the action sets for the decision maker and opponent, respectively. The process then operates as follows. In a state $x_n$, $n = 0, 1, \ldots$, the controller chooses an action $a_n \in A(x_n)$, while the opponent selects $\mu_n(\cdot)[x_n, a_n] \in \Pr(X)$. As a consequence, the controller pays $\gamma c(x_n, a_n) - R(\mu_n \| q(\cdot | x_n, a_n))$ to his opponent, and the system moves to the next state according to the probability distribution $\mu_n(\cdot)[x_n, a_n]$.

We shall deal with the following classes of strategies. It will cause no confusion if we continue to use the same letters to denote strategies for the controller. Namely, $\pi$ stands for a randomized control strategy (policy), whereas $f$ denotes a stationary strategy. We write $\Pi$ and $F$ to denote the sets of corresponding strategies. For the opponent's class of strategies, we confine to the stationary one, which is identified with the class $P$ of stochastic kernels $p$ on $X$ given $K$.

Let $(\Omega, \mathcal{F})$ be the measurable space consisting of the sample space $\Omega = (X \times A)^\infty$ and its product $\sigma$-algebra $\mathcal{F}$. Then for an initial state $x \in X$, and strategies $\pi$ and $p$, there exists a unique probability measure $\mathcal{P}_x^{\pi p}$ and, again, a stochastic process $\{(x_k, a_k)\}$ is defined $(\Omega, \mathcal{F})$ in a canonical way, where $x_k$ denotes the state at time $k$ and $a_k$ is the action for the controller. With some abuse of notation, we let $h_k$ stand for the history of the process up to the $k$th state, that is,

$$h_k = (x_0, a_0, x_1, \ldots, a_{k-1}, x_k).$$

The corresponding expectation operator is denoted by $\mathcal{E}_x^{\pi p}$.

For fixed $x \in X$, $\pi \in \Pi$ and $p \in P$, we define the following functional costs:

$$(1) \quad V_\beta(x, \pi, p) = \sum_{k=0}^\infty \beta^k \mathcal{E}_x^{\pi p}[\gamma c(x_k, a_k) - R(p(\cdot | x_k, a_k) \| q(\cdot | x_k, a_k))],$$

where $\beta \in (0, 1)$ is the discount factor, and

$$j(x, \pi, p) = \limsup_{n \to \infty} \frac{1}{n\gamma} \sum_{k=0}^{n-1} \mathcal{E}_x^{\pi p}[\gamma c(x_k, a_k) - R(p(\cdot | x_k, a_k) \| q(\cdot | x_k, a_k))].$$

Note that, since the function $R(\cdot \| \cdot)$ is lower semicontinuous on $\Pr(X) \times \Pr(X)$ and $p$ and $q$ are stochastic kernels [i.e., measurable functions of $(x, a)$], it follows that the mapping

$$(x, a) \mapsto R(p(\cdot | x, a) \| q(\cdot | x, a))$$

is measurable (Lemma 1.4.3(f) in [12]). Observe that $V_\beta(x, \pi, p)$ and $j(x, \pi, p)$ might be undetermined, because $c$ can be unbounded. We thus restrict the set of admissible strategies for the opponent in the following way.

DEFINITION 1. Given $\pi = \{\pi_k\} \in \Pi$, we say that $p \in P$ is a $\pi$-admissible strategy iff

$$\int_{A(x_k)} R(p(\cdot|x_k,a)\|q(\cdot|x_k,a))\pi_k(da|h_k) < +\infty, \tag{2}$$

and moreover, there exists a constant $C \geq 0$, possibly depending on $\pi$ and $p$, such that

$$\int_{A(x_k)} [\gamma c(x_k,a) - R(p(\cdot|x_k,a)\|q(\cdot|x_k,a))]\pi_k(da|h_k) + C \geq 0,$$

for all histories of the process $h_k$, $k \geq 0$, induced by $p$ and $\pi$. We denote this set by $Q(\pi)$. [Note that this set is nonempty, since $p = q \in Q(\pi)$ for any $\pi \in \Pi$.]

Let us introduce the following notation. For any $\pi \in \Pi$, $p \in Q(\pi)$ and $n \geq 1$, define

$$J_n(x,\pi) = \log E_x^\pi \exp\left\{\gamma \sum_{k=0}^{n-1} c(x_k,a_k)\right\}, \tag{3}$$

and

$$j_n(x,\pi,p) = \sum_{k=0}^{n-1} \mathcal{E}_x^{\pi p}[\gamma c(x_k,a_k) - R(p(\cdot|x_k,a_k)\|q(\cdot|x_k,a_k))].$$

Now we are ready to present the result that was originally proved in [16] for Markov strategies. However, it still remains valid when arbitrary strategies for the decision maker are considered. Therefore, for the sake of clarity, we state the result with its proof.

PROPOSITION 1. *Let $x \in X$ and $p \in Q(\pi)$. Then:*

(a) $\sup_{p \in Q(\pi)} j_n(x,\pi,p) \leq J_n(x,\pi)$ *for each* $n \geq 1$,
(b) $\limsup_{n \to \infty} \sup_{p \in Q(\pi)} \frac{1}{n} j_n(x,\pi,p) \leq \gamma J(x,\pi)$.

PROOF. (a) Let $p \in Q(\pi)$ be any stochastic kernel. For $n = 1$, we conclude

$$j_1(x,\pi,p) \leq \mathcal{E}_x^{\pi p}(\gamma c(x,a_0)) \leq \log E_x^\pi e^{\gamma c(x,a_0)} = J_1(x,\pi),$$

where the first inequality holds since the relative entropy is nonnegative, and the second one is due to Jensen's inequality. Now assume that the hypothesis is true for some $n \geq 1$. Clearly,

$$j_{n+1}(x,\pi,p) = \sum_{k=0}^{n} \mathcal{E}_x^{\pi p}[\gamma c(x_k,a_k) - R(p(\cdot|x_k,a_k)\|q(\cdot|x_k,a_k))]$$

$$= \mathcal{E}_x^{\pi p} \sum_{k=0}^{n} [\gamma c(x_k,a_k) - R(p(\cdot|x_k,a_k)\|q(\cdot|x_k,a_k))], \qquad n \geq 1.$$

Denote by $\pi^{(1)}$ the "1-shifted" strategy, that is,

$$\pi_k^{(1)}(\cdot|h_k) = \pi_{k+1}(\cdot|x_0, a_0, h_k), \qquad k \geq 0.$$

Then, we have

$$\begin{aligned}
&j_{n+1}(x, \pi, p) \\
&= \mathcal{E}_x^{\pi p}[\gamma c(x, a_0) + j_n(x_1, \pi^{(1)}, p) - R(p(\cdot|x, a_0)\|q(\cdot|x, a_0))] \\
&\leq \mathcal{E}_x^{\pi p}(\gamma c(x, a_0)) \\
&\quad + \mathcal{E}_x^{\pi p}(\mathcal{E}_x^{\pi p}\{[J_n(x_1, \pi^{(1)}) - R(p(\cdot|x, a_0)\|q(\cdot|x, a_0))]|a_0\}) \\
&= E_x^\pi \log e^{\gamma c(x, a_0)} \\
&\quad + \mathcal{E}_x^{\pi p}\left[\int_X J_n(x_1, \pi^{(1)}) p(dx_1|x, a_0) - R(p(\cdot|x, a_0)\|q(\cdot|x, a_0))\right] \\
&\leq \int_{A(x)} \log e^{\gamma c(x, a_0)} \pi_0(da_0|x) \\
&\quad + \int_{A(x)} \log \int_X e^{J_n(x_1, \pi^{(1)})} q(dx_1|x, a_0) \pi_0(da_0|x) \\
&= \int_{A(x)} \log \int_X E_{x_1}^{\pi^{(1)}} e^{\gamma c(x, a_0) + \sum_{k=1}^{n+1} \gamma c(x_k, a_k)} q(dx_1|x, a_0) \pi_0(da_0|x) \\
&\leq \log \int_{A(x)} \int_X E_{x_1}^{\pi^{(1)}} e^{\gamma c(x, a_0) + \sum_{k=1}^{n+1} \gamma c(x_k, a_k)} q(dx_1|x, a_0) \pi_0(da_0|x) \\
&= J_{n+1}(x, \pi).
\end{aligned}$$

Clearly, the first inequality follows from the induction hypothesis. The third inequality is due to Jensen's inequality, whilst the second one follows from Lemma A in the Appendix. Since $p \in Q(\pi)$ is arbitrary, we get the desired conclusion.

Part (b) follows directly from part (a). □

REMARK 2. Note that in the proof of Proposition 1 we did not really have to use the fact that $p \in Q(\pi)$. The only assumption which plays an essential role is condition (2). Namely, it guarantees that $j_n(x, \pi, p)$ is well defined for all $n \geq 1$, $x \in X$ and $\pi \in \Pi$. However, in Definition 1 we restrict the opponent's class of strategies to the set $Q(\pi)$ in order to be able to apply the Hardy–Littlewood theorem. In actual fact, later on it will be clear that the set $Q(\pi)$, where $\pi \in \Pi$, is sufficiently large. Namely, the supremum of certain discounted functional costs over the set $Q(\pi)$ will not change if we add new elements to $Q(\pi)$; see the proofs of Lemmas 1 and 2.

Let $\widetilde{\pi}$ be as in assumption (G) and let $p \in Q(\widetilde{\pi})$. Then from the Hardy–Littlewood theorem (Theorem H.2 in [13]), we get

$$\limsup_{\beta \to 1}(1-\beta)V_\beta(x,\widetilde{\pi},p) \leq \limsup_{n \to \infty} \frac{1}{n} j_n(x,\widetilde{\pi},p)$$

and from Proposition 1(b),

$$\limsup_{n \to \infty} \sup_{p \in Q(\widetilde{\pi})} \frac{1}{n} j_n(x,\widetilde{\pi},p) \leq \gamma J(x,\widetilde{\pi}).$$

Combining these two inequalities, we conclude that

$$\limsup_{\beta \to 1}(1-\beta)V_\beta(x,\widetilde{\pi},p) \leq \gamma J(x,\widetilde{\pi}) \qquad \text{for every } p \in Q(\widetilde{\pi}).$$

This in turn yields

(4) $$\limsup_{\beta \to 1}(1-\beta)V_\beta(x) \leq \gamma J(x,\widetilde{\pi}),$$

where $V_\beta(x)$ is the upper value of functional cost (1), that is,

$$V_\beta(x) = \inf_{\pi \in \Pi} \sup_{p \in Q(\pi)} V_\beta(x,\pi,p).$$

Consequently, inequality (4) and assumption (G) together lead to the following:

(5) $$V_\beta(x) < +\infty$$

for each $x \in X$ and $\beta \in (0,1)$. In addition, $V_\beta(x) \geq 0$. Now defining

$$\rho := \inf_{x \in X} \inf_{\pi \in \Pi} J(x,\pi), \qquad m_\beta := \inf_{x \in X} V_\beta(x)$$

and observing that

(6) $$\limsup_{\beta \to 1}(1-\beta)m_\beta \leq \gamma\rho,$$

one can deduce that there exists a sequence of discount factors $\{\beta_n\}$ converging to 1 for which

(7) $$\lim_{n \to \infty}(1-\beta_n)m_{\beta_n} = l,$$

where $l$ is a certain nonnegative constant.

**3. A solution to the auxiliary discounted minimax problem.** The main thrust of this section is to solve the auxiliary discounted minimax problem introduced in the previous section. In other words, we look for a discounted functional equation whose solution is the function $V_\beta$. This is done by an approximation of the above-mentioned minimax models by ones with bounded cost functions. These models in turn are solved by a fixed point argument in Proposition 1. Next, we show in Lemma 1 that the corresponding solutions equal the upper values of some discounted costs on the infinite horizon. Finally, the limit passage in Lemma 2 gives the desired discounted functional equation with the function $V_\beta$ as a solution.

We shall need the following two sets of compactness-semicontinuity assumptions, which will be used alternatively.

CONDITION (**S**).

(i) The set $A(x)$ is compact.
(ii) For each $x \in X$ and every Borel set $D \subset X$, the function $q(D|x,\cdot)$ is continuous on $A(x)$.
(iii) The cost function $c(x,\cdot)$ is lower semicontinuous for each $x \in X$.

CONDITION (**W**).

(i) The set $A(x)$ is compact and the set-valued mapping $x \mapsto A(x)$ is upper semicontinuous, that is, $\{x \in X : A(x) \cap B \neq \varnothing\}$ is closed for every closed set $B$ in $A$.
(ii) The transition law $q$ is weakly continuous on $K$, that is, the function

$$(x,a) \mapsto \int_X u(y)q(dy|x,a), \qquad (x,a) \in K,$$

is continuous function for each bounded continuous function $u$.
(iii) The cost function $c$ is lower semicontinuous on $K$.

By $L_b(X)$ and $B_b(X)$, we denote the set of all bounded lower semicontinuous and bounded Borel measurable functions on $X$, respectively. Further, let $\mathbb{N}$ stand for the set of positive integers. Choose $N \in \mathbb{N}$ and define the truncated cost function

$$c^N(x,a) = \min\{N, c(x,a)\}.$$

The following result was proved under Condition (**W**) for bounded cost functions by a fixed point argument; see page 72 in [10]. However, a simple and obvious modification of the proof gives the conclusion under Condition (**S**) as well.

PROPOSITION 2. *Under* (**W**) [(**S**)], *for any discount factor $\beta \in (0,1)$ and a number $N \in \mathbb{N}$, there exists a unique function $w_\beta^N \in L_b(X)$ [$w_\beta^N \in B_b(X)$] such that*

$$(8) \qquad e^{w_\beta^N(x)} = \min_{a \in A(x)} \left[ e^{\gamma c^N(x,a)} \int_X e^{\beta w_\beta^N(y)} q(dy|x,a) \right]$$

*for each $x \in X$, and*

$$(9) \qquad 0 \leq (1-\beta) w_\beta^N(x) \leq N\gamma.$$

*Moreover, there exists a stationary strategy $f^0 \in F$ (possibly depending on $\beta$ and $N$) that attains the minimum in (8).*

Let $\beta$ and $N$ be fixed just in the next lemma.

LEMMA 1. *Assume* (**W**) *or* (**S**). *Then, it holds*

$$(10) \qquad w_\beta^N(x) = \inf_{\pi \in \Pi} \sup_{p \in Q(\pi)} \sum_{k=0}^\infty \mathcal{E}_x^{\pi p} \beta^k [\gamma c^N(x_k, a_k) - R(p(\cdot|x_k, a_k) \| q(\cdot|x_k, a_k))]$$

*for any initial state $x \in X$.*

PROOF. Note that (8) can be rewritten in the following equivalent form:

$$(11) \qquad w_\beta^N(x) = \min_{a \in A(x)} \left[ \gamma c^N(x,a) + \log \int_X e^{\beta w_\beta^N(y)} q(dy|x,a) \right].$$

Applying Lemma A in the Appendix to (11), we get

$$(12) \qquad w_\beta^N(x) = \min_{a \in A(x)} \sup_{\mu \in \Delta(x,a)} \left[ \gamma c^N(x,a) - R(\mu \| q(\cdot|x,a)) + \beta \int_X w_\beta^N(y) \mu(dy) \right],$$

with

$$\Delta(x,a) := \{ \mu \in \Pr(X) : R(\mu \| q(\cdot|x,a)) < +\infty \}, \qquad (x,a) \in K.$$

Moreover, the measure

$$\mu^0(dy)[x,a] = \frac{e^{\beta w_\beta^N(y)} q(dy|x,a)}{\int_X e^{\beta w_\beta^N(y)} q(dy|x,a)}$$

achieves the supremum in (12). Put

$$(13) \qquad p^0(dy|x,a) = \mu^0(dy)[x,a] \qquad \text{for each } (x,a) \in K.$$

Note that $p^0 \in Q(\pi)$ for any strategy $\pi \in \Pi$. This directly follows from the definition of $R(p^0(\cdot|x,a)\|q(\cdot|x,a))$ and (9). Simple calculations give the upper bound

$$R(p^0(\cdot|x,a)\|q(\cdot|x,a)) \leq 2\frac{\beta N\gamma}{1-\beta}\exp\left(\frac{\beta N\gamma}{1-\beta}\right)$$

for every $(x,a) \in K$.

Let $p^0$ be defined as in (13). By (12), we then have

$$w_\beta^N(x) \leq \gamma c^N(x,a) - R(p^0(\cdot|x,a)\|q(\cdot|x,a)) + \beta\int_X w_\beta^N(y)p^0(dy|x,a).$$

By iteration of this inequality $n$ times, it follows

$$w_\beta^N(x) \leq \sum_{k=0}^n \beta^k \mathcal{E}_x^{\pi p^0}[\gamma c^N(x_k,a_k) - R(p^0(\cdot|x_k,a_k)\|q(\cdot|x_k,a_k))]$$
$$+ \beta^{n+1} \mathcal{E}_x^{\pi p^0} w_\beta^N(x_{n+1}),$$

where $\pi$ is any strategy for the controller. Now, letting $n \to \infty$ and making use of (9), we conclude

$$w_\beta^N(x) \leq \sum_{k=0}^\infty \beta^k \mathcal{E}_x^{\pi p^0}[\gamma c^N(x_k,a_k) - R(p^0(\cdot|x_k,a_k)\|q(\cdot|x_k,a_k))].$$

Since $\pi$ is arbitrary, we get

$$w_\beta^N(x) \leq \inf_{\pi \in \Pi} \sum_{k=0}^\infty \beta^k \mathcal{E}_x^{\pi p^0}[\gamma c^N(x_k,a_k) - R(p^0(\cdot|x_k,a_k)\|q(\cdot|x_k,a_k))]$$

(14)
$$\leq \inf_{\pi \in \Pi} \sup_{p \in Q(\pi)} \sum_{k=0}^\infty \beta^k \mathcal{E}_x^{\pi p}[\gamma c^N(x_k,a_k)$$
$$- R(p(\cdot|x_k,a_k)\|q(\cdot|x_k,a_k))].$$

Note that inequality (14) is valid because $p^0 \in Q(\pi)$.

On the other hand, by (12), we can write

$$w_\beta^N(x) \geq \gamma c^N(x,f^0(x)) - R(p(\cdot|x,f^0(x))\|q(\cdot|x,f^0(x)))$$
$$+ \beta \int_X w_\beta^N(y)p(dy|x,f^0(x)),$$

with $f^0$ as in Proposition 2 and any $p \in Q(f^0)$. Proceeding along the same line, we infer

$$w_\beta^N(x) \geq \sum_{k=0}^\infty \beta^k \mathcal{E}_x^{f^0 p}[\gamma c^N(x_k,f^0(x_k)) - R(p(\cdot|x_k,f^0(x_k))\|q(\cdot|x_k,f^0(x_k)))].$$

Since $p \in Q(f^0)$ is arbitrary, we easily deduce

$$w_\beta^N(x) \geq \sup_{p \in Q(f^0)} \sum_{k=0}^{\infty} \beta^k \mathcal{E}_x^{f^0 p}[\gamma c^N(x_k, f^0(x_k))$$
$$- R(p(\cdot|x_k, f^0(x_k))\|q(\cdot|x_k, f^0(x_k)))]$$

(15)

$$\geq \inf_{\pi \in \Pi} \sup_{p \in Q(\pi)} \sum_{k=0}^{\infty} \beta^k \mathcal{E}_x^{\pi p}[\gamma c^N(x_k, a_k)$$
$$- R(p(\cdot|x_k, a_k)\|q(\cdot|x_k, a_k))].$$

Finally, combining (14) with (15) completes the proof. $\square$

In the remainder of the paper, we shall use the following notation. Let $L(X)$ denote the set of all lower semicontinuous functions on $X$, whereas $B(X)$ stands for the set of all Borel measurable functions on $X$.

LEMMA 2. *Let* (W) [(S)] *hold and* $\beta \in (0,1)$. *Then, we have the following:*

(a) *The function*

$$w_\beta(x) := \lim_{N \to \infty} w_\beta^N(x)$$

*is finite and nonnegative for each* $x \in X$. *Moreover,* $w_\beta \in L(X)$ [$w_\beta \in B(X)$].

(b) *The functional equation holds*

(16) $$e^{w_\beta(x)} = \min_{a \in A(x)} \left[ e^{\gamma c(x,a)} \int_X e^{\beta w_\beta(y)} q(dy|x,a) \right]$$

*for all* $x \in X$. *Furthermore, there exists a Borel measurable selector* $f_\beta \in F$ *of the minima in* (16).

(c) *For any* $x \in X$, $w_\beta(x) = V_\beta(x)$.

PROOF. Let $x \in X$ and $\beta \in (0,1)$ be fixed. From (10), it is easily seen that the sequence $\{w_\beta^N(x)\}$ is nondecreasing in $N$. Therefore, $w_\beta(x) = \lim_{N \to \infty} w_\beta^N(x)$ exists and by (9), it is nonnegative. Clearly, under (S), $w_\beta \in B(X)$, whereas, under (W), $w_\beta \in L(X)$; see Proposition 10.1 in [26].

In order to prove that $w_\beta(x)$ is finite for each $x \in X$, observe first that, for any $\pi \in \Pi$, $p \in Q(\pi)$ and $N \in \mathbb{N}$,

$$V_\beta(x, \pi, p) = \sum_{k=0}^{\infty} \beta^k \mathcal{E}_x^{\pi p}[\gamma c(x_k, a_k) - R(p(\cdot|x_k, a_k)\|q(\cdot|x_k, a_k))]$$
$$\geq \sum_{k=0}^{\infty} \beta^k \mathcal{E}_x^{\pi p}[\gamma c^N(x_k, a_k) - R(p(\cdot|x_k, a_k)\|q(\cdot|x_k, a_k))].$$

Moreover, from Lemma 1, we have

$$V_\beta(x) = \inf_{\pi \in \Pi} \sup_{p \in Q(\pi)} V_\beta(x, \pi, p)$$

$$\geq \inf_{\pi \in \Pi} \sup_{p \in Q(\pi)} \sum_{k=0}^{\infty} \beta^k \mathcal{E}_x^{\pi p}[\gamma c^N(x_k, a_k) - R(p(\cdot|x_k, a_k)\|q(\cdot|x_k, a_k))]$$

$$= w_\beta^N(x).$$

Hence, letting $N \to \infty$, it follows

(17) $$V_\beta(x) \geq \lim_{N \to \infty} w_\beta^N(x) = w_\beta(x).$$

By (5), $V_\beta(x)$ is finite for each $x \in X$, so is $w_\beta(x)$. This finishes the proof of part (a).

In order to prove part (b), note that by (11) and part (a) the limit

(18) $$\lim_{N \to \infty} \min_{a \in A(x)} \left[ \gamma c^N(x, a) + \log \int_X e^{\beta w_\beta^N(y)} q(dy|x, a) \right]$$

exists. Since the first and the second term in (18) are nondecreasing and (**W**) or (**S**) holds, then we may interchange the limit with the minimum (see Proposition 10.1 in [26]). Furthermore, making use of the Lebesgue monotone convergence theorem, we conclude (16). The existence of a Borel measurable selector $f_\beta \in F$ follows from the compactness–semicontinuity assumptions and Proposition D.5 in [17].

We now turn to proving part (c). Again, taking a logarithm on both sides of (16), it follows

(19) $$w_\beta(x) = \min_{a \in A(x)} \left[ \gamma c(x, a) + \log \int_X e^{\beta w_\beta(y)} q(dy|x, a) \right].$$

Applying Lemma A in the Appendix to (19), we easily obtain

(20)
$$w_\beta(x)$$
$$= \min_{a \in A(x)} \sup_{\mu \in \Delta(x,a)} \left[ \gamma c(x, a) - R(\mu\|q(\cdot|x, a)) + \beta \int_X w_\beta(y) \mu(dy) \right],$$

with

$$\Delta(x, a) = \{\mu \in \Pr(X) : R(\mu\|q(\cdot|x, a)) < +\infty\}, \qquad (x, a) \in K.$$

Observe that by (20), for any $p \in Q(f_\beta)$, the following holds:

$$w_\beta(x) \geq \gamma c(x, f_\beta(x)) - R(p(\cdot|x, f_\beta(x))\|q(\cdot|x, f_\beta(x)))$$
$$+ \beta \int_X w_\beta(y) p(dy|x, f_\beta(x)).$$

Iterating this inequality $n$ times, we immediately obtain

$$w_\beta(x) \geq \sum_{k=0}^n \beta^k \mathcal{E}_x^{f_\beta p}[\gamma c(x_k, f_\beta(x_k))$$
$$- R(p(\cdot|x_k, f_\beta(x_k))\|q(\cdot|x_k, f_\beta(x_k)))]$$
$$(21) \qquad + \beta^{n+1} \mathcal{E}_x^{f_\beta p} w_\beta(x_{n+1})$$
$$\geq \sum_{k=0}^n \beta^k \mathcal{E}_x^{f_\beta p}[\gamma c(x_k, f_\beta(x_k))$$
$$- R(p(\cdot|x_k, f_\beta(x_k))\|q(\cdot|x_k, f_\beta(x_k)))].$$

Now note that, by Definition 1,

$$\mathcal{E}_x^{f_\beta p}[\gamma c(x_k, f_\beta(x_k)) - R(p(\cdot|x_k, f_\beta(x_k))\|q(\cdot|x_k, f_\beta(x_k)))] \geq -C,$$

for some $C \geq 0$ and $k \geq 1$. Thus, letting $n \to \infty$ in (21), it follows

$$w_\beta(x) \geq \sum_{k=0}^\infty \beta^k \mathcal{E}_x^{f_\beta p}[\gamma c(x_k, f_\beta(x_k)) - R(p(\cdot|x_k, f_\beta(x_k))\|q(\cdot|x_k, f_\beta(x_k)))]$$
$$= V_\beta(x, f_\beta, p).$$

Since $p \in Q(f_\beta)$ is arbitrary, we see that

$$(22) \qquad w_\beta(x) \geq \sup_{p \in Q(f_\beta)} V_\beta(x, f_\beta, p) \geq V_\beta(x).$$

Inequalities (17) and (22) combined conclude the proof of part (c). □

**4. A solution to the risk-sensitive control problem.** For any $x \in X$ and any discount factor $\beta \in (0,1)$, define

$$h_\beta(x) := V_\beta(x) - m_\beta$$

with $m_\beta = \inf_{x \in X} V_\beta(x)$. Obviously, $h_\beta$ is nonnegative.

The following boundedness assumption is supposed to hold true. As mentioned in the Introduction, we put off discussing it until Section 5:

CONDITION (**B**). For any $x \in X$, $\sup_{\beta \in (0,1)} h_\beta(x) < +\infty$.

REMARK 3. A similar assumption and its equivalent variants were used to study the expected average cost criterion for Markov decision processes in the risk-neutral setting [17, 27, 28]. Roughly speaking, Hernández-Lerma and Lasserre [17], Schäl [27], and Sennott [28] assume that the family of the so-called normalized $\beta$-discounted cost functions is bounded. This assumption, however, simply holds for ergodic Markov decision processes. More precisely, if the $n$-step transition probabilities converge to the unique invariant probability measure geometrically fast, and the cost functions are bounded (or more generally satisfy a certain growth hypothesis), then the aforementioned family of functions is pointwise relatively compact [21, 22]. It is worth pointing out that this requirement is crucial to obtain the optimality inequality in the risk-neutral case; see [27, 28]. In Section 5 we provide an example that illustrates that also in the risk-sensitive case Condition (**B**) cannot be weakened.

We shall need the following two versions of Fatou's lemma for converging measures.

LEMMA 3. *Let $\{\mu_n\}$ be a sequence of probability measures converging to $\mu \in \Pr(X)$ and let $\{h_n\}$ be a sequence of measurable nonnegative functions on $X$. Then,*

$$\int_X h(y)\mu(dy) \leq \liminf_{n\to\infty} \int_X h_n(y)\mu_n(dy)$$

*in the following cases:*

(a) $\{\mu_n\}$ *converges setwise to $\mu$ [i.e., $\int_X f(y)\,d\mu_n(y) \to \int_X f(y)\,d\mu(y)\,\forall f \in B_b(X)$], and $h(x) = \liminf_{n\to\infty} h_n(x)$;*

(b) $\{\mu_n\}$ *converges weakly to $\mu$, and $h(x) = \inf\{\liminf_{n\to\infty} h_n(x_n): x_n \to x\}$; moreover, $h \in L(X)$.*

PROOF. Part (a) is due to Royden [25], page 231, whereas part (b) was proved by Serfozo [29]. For the proof of lower semicontinuity of $h$, the reader is referred to Lemma 3.1 in [22]. □

Now we are in a position to state the main result of the paper. This theorem concerns a study of the risk-sensitive average cost optimality inequality, which is sufficient to establish the existence of an optimal stationary policy.

THEOREM 1. *Assume* (**B**) *and* (**W**) *[or* (**S**)*]. Then, for each risk factor $\gamma > 0$, there exist a constant $\widehat{l}$ and a nonnegative function $h \in L(X)$ [$h \in B(X)$] and $\widehat{f} \in F$ such that*

$$
\begin{aligned}
h(x) + \widehat{l} &\geq \min_{a \in A(x)} \left[ \gamma c(x,a) + \log \int_X e^{h(y)} q(dy|x,a) \right] \\
&= \gamma c(x, \widehat{f}(x)) + \log \int_X e^{h(y)} q(dy|x, \widehat{f}(x))
\end{aligned}
\tag{23}
$$

*for all $x \in X$. Moreover,*

$$\frac{\widehat{l}}{\gamma} = \inf_{\pi \in \Pi} J(x, \pi) = J(x, \widehat{f}).$$

*In other words, $\widehat{l}/\gamma$ is the optimal risk-sensitive average cost and $\widehat{f}$ is a risk-sensitive average cost optimal stationary policy.*

REMARK 4. (a) There are two papers [16, 27] that can be treated as predecessors of our work. They both deal with the optimality inequality but within two different frameworks. The first work [16] establishes the optimality equation for the risk-sensitive dynamic programming on a denumarable state space. In the other one, the result is obtained for Markov control processes on an uncountable state space for the risk factor $\gamma = 0$. From this point of view, our result is an extention of Theorem 4.1 in [16] to a general state space and Theorem 3.8 in [27] to the risk-sensitive case. Moreover, the common feature of the discussed results is that their proofs are based on the vanishing discount factor approach. Our proof also relies on this method, and similarly, as in [27] or [21, 22], makes use of the Fatou lemmas for setwise and weakly convergent measures.

(b) Finally, it is also worth mentioning that there are papers studying the *optimality equation* in the risk-sensitive dynamic programming, which is of the following form:

(24) $$h(x) + \widehat{l} = \min_{a \in A(x)} \left[ \gamma c(x, a) + \log \int_X e^{h(y)} q(dy|x, a) \right].$$

The constant $\frac{\widehat{l}}{\gamma}$ is (under suitable assumptions) an optimal cost with respect to the risk-sensitive average cost criterion. Let us mention and discuss a few representative papers that deal with equation (24). In [8, 15] Markov control models satisfying a simultaneous Doeblin condition, on a finite and countable state space, respectively, are considered. The cost functions are supposed to be bounded and the risk factor must be sufficiently small. Otherwise, as argued in [8], the optimality equation need not have a solution.

In [10] Di Masi and Stettner extend the result to a general state space by retaining bounded cost functions and replacing a simultaneous Doeblin condition with a very strong assumption on transition probabilities. In [11], however, they replace this assumption by one imposed on the risk coefficient. Finally, the class of Markov control models that requires neither any ergodicity conditions nor the smallness of the risk factor was pointed out by Jaśkiewicz in [20].

Fairly recently Borkar and Meyn [5] considered Markov decision processes with unbounded cost functions on a denumarable state space. Their result assumes the following: the state space is irreducible under all Markov policies, the costs are norm-like, and there exists a policy that induces a finite average risk-sensitive cost. Moreover, their proof is based on a multiplicative ergodic theorem that was studied in more detail in [1].

PROOF OF THEOREM 1. Let $\{\beta_n\}$ be a sequence of discount factors converging to 1 for which (7) holds. Defining
$$\widehat{l} := l = \lim_{n\to\infty} (1-\beta_n) m_{\beta_n}$$
and applying (6), we note that
$$(25) \qquad \frac{\widehat{l}}{\gamma} \leq \inf_{\pi\in\Pi} J(x,\pi)$$
for any $x \in X$. Assume for a while that inequality (23) is satisfied and there exists $\widehat{f} \in F$ as in the statement of Theorem 1. We prove that $\widehat{f}$ is an optimal policy. From (23), we have
$$h(x) \geq \gamma c(x, \widehat{f}(x)) - \widehat{l} + \log \int_X e^{h(y)} q(dy|x, \widehat{f}(x)).$$
By iteration of this inequality $n$ times, we obtain
$$h(x) \geq \log E_x^\pi \exp\left(\sum_{k=0}^n \gamma c(x_k, \widehat{f}(x_k)) + h(x_{n+1})\right) - (n+1)\widehat{l}.$$
Since $h$ is nonnegative, we infer
$$\frac{h(x)}{n+1} + \widehat{l} \geq \frac{J_{n+1}(x, \widehat{f})}{n+1},$$
with $J_{n+1}(x, \widehat{f})$ defined in (3). Letting $n \to \infty$, it follows
$$(26) \qquad \frac{\widehat{l}}{\gamma} \geq J(x, \widehat{f}), \qquad x \in X.$$
Hence, (25) and (26) together imply
$$\frac{\widehat{l}}{\gamma} = J(x, \widehat{f}) = \inf_{\pi\in\Pi} J(x, \Pi)$$
for each $x \in X$.

We next focus on showing inequality (23). Let $n \geq 1$ and put $h_n := h_{\beta_n}$, $f_n := f_{\beta_n}$. Note that (19) can be rewritten in the following form:
$$(27) \quad \begin{aligned} (1-\beta_n) m_{\beta_n} + h_n(x) &= \min_{a\in A(x)} \left[\gamma c(x,a) + \log \int_X e^{\beta_n h_n(y)} q(dy|x,a)\right] \\ &= \gamma c(x, f_n(x)) + \log \int_X e^{\beta_n h_n(y)} q(dy|x, f_n(x)). \end{aligned}$$

(i) Assume first (**S**) and define
$$h(x) = \liminf_{n\to\infty} h_n(x).$$
Taking the lim inf on both sides of (27), we get
$$\liminf_{n\to\infty}((1-\beta_n)m_{\beta_n} + h_n(x))$$
$$= \widehat{l} + h(x) = \liminf_{n\to\infty} \min_{a\in A(x)}\left[\gamma c(x,a) + \log\int_X e^{\beta_n h_n(y)}q(dy|x,a)\right].$$
Making use of Lemma 3(a) and the measurable selection theorem (see Proposition D.5(a) in [17]), one can prove that there exists $\widehat{f} \in F$ such that (23) holds.

(ii) Now assume (**W**). Fix $x_0 \in X$ and choose any $x_n \to x_0$, $n \to \infty$. Take a subsequence $\{n_k\}$ of positive integers such that
$$\liminf_{n\to\infty} h_n(x_n) = \lim_{k\to\infty} h_{n_k}(x_{n_k}).$$
Then by (27),
$$\liminf_{n\to\infty}((1-\beta_n)m_{\beta_n} + h_n(x_n))$$
$$= \widehat{l} + \liminf_{n\to\infty} h_n(x_n) = \widehat{l} + \lim_{k\to\infty} h_{n_k}(x_{n_k})$$
(28)
$$= \lim_{k\to\infty} \min_{a\in A(x_{n_k})}\left[\gamma c(x_{n_k},a) + \log\int_X e^{\beta_{n_k} h_{n_k}(y)}q(dy|x_{n_k},a)\right]$$
$$= \lim_{k\to\infty}\left[\gamma c(x_{n_k}, f_{n_k}(x_{n_k})) + \log\int_X e^{\beta_{n_k} h_{n_k}(y)}q(dy|x_{n_k}, f_{n_k}(x_{n_k}))\right].$$
Note that $G = \{x_0\} \cup \{x_n\}$ is compact in $X$. From the upper semicontinuity of $x \mapsto A(x)$, compactness of every $A(z)$ and Berge's theorem (see [2] or Theorem 7.4.2 in [23]), it follows that $\bigcup_{z\in G} A(z)$ is compact in $A$. Therefore, $\{f_{n_k}(x_{n_k})\}$ has a subsequence converging to some $a_0 \in A$. By (**W**)(i), $a_0 \in A(x_0)$, that is, $(x_0, a_0) \in K$. Without loss of generality, assume that $f_{n_k}(x_{n_k}) \to a_0$, $k \to \infty$. By the lower semicontinuity of the cost function $c$ and (28), we have
$$\widehat{l} + \liminf_{n\to\infty} h_n(x_n) \geq \gamma c(x_0, a_0) + \lim_{k\to\infty} \log\int_X e^{\beta_{n_k} h_{n_k}(y)}q(dy|x_{n_k}, f_{n_k}(x_{n_k})).$$
This and Lemma 3(b) imply that
$$\widehat{l} + \liminf_{n\to\infty} h_n(x_n) \geq \gamma c(x_0, a_0) + \log\int_X e^{\widetilde{h}(y)}q(dy|x_0, a_0),$$
where $e^{\widetilde{h}}$ is the generalized lim inf of the sequence $e^{\widetilde{h}_k} = e^{h_{n_k}}$. Clearly, $h \leq \widetilde{h}$. By Lemma 3(b), $h \in L(X)$. Thus,
(29) $$\widehat{l} + \liminf_{n\to\infty} h_n(x_n) \geq \gamma c(x_0, a_0) + \log\int_X e^{h(y)}q(dy|x_0, a_0).$$

Since $x_n \to x_0$ was chosen arbitrarily, we infer from (29) that

$$\widehat{l} + h(x_0) \geq \gamma c(x_0, a_0) + \log \int_X e^{h(y)} q(dy|x_0, a_0).$$

The last inequality shows that, for any $x \in X$, there exists an $a_x \in A(x)$ such that

(30)
$$\widehat{l} + h(x) \geq \gamma c(x, a_x) + \log \int_X e^{h(y)} q(dy|x, a_x)$$
$$\geq \min_{a \in A(x)} \left[ \gamma c(x, a) + \int_X e^{h(y)}(y) q(dy|x, a) \right].$$

By our compactness–semicontinuity assumptions and Proposition D.5(b) in [17], there exists some $\widehat{f} \in F$ such that (23) holds. □

**5. A discussion.** This section is devoted to a discussion of Condition (**B**). We start with revisiting Example 3.1 in [8].

EXAMPLE 1. Put $X = \{0, 1\}$, $A = \{a\}$, $c(x) := c(x, a) = x$ and the transition matrix is as follows:

$$\begin{bmatrix} 1 & 0 \\ \rho & 1 - \rho \end{bmatrix},$$

where $\rho \in (0, 1)$. Recall that the following was proved.

Let us consider three cases for the risk factor $\gamma$:

(I) $\gamma < -\log(1 - \rho)$,
(II) $\gamma = -\log(1 - \rho)$,
(III) $\gamma > -\log(1 - \rho)$.

Then if (I) or (II) hold, the optimal risk-sensitive average cost equals 0 and is independent of the initial state. In case (III) we have $J^*(0) = 0$ and $J^*(1) = 1 + \frac{\log(1-\rho)}{\gamma} > 0$. In addition, it is interesting to observe that, for (II) and (III) cases, there does not exist a function $h: X \mapsto \mathbb{R}$ such that optimality inequality (23) is satisfied. Indeed, to see this take $x = 1$ and consider (III). The optimality inequality is then as follows:

$$\gamma J^*(1) + h(1) = \gamma + \log(1 - \rho) + h(1) \geq \gamma + \log(e^{h(1)}(1 - \rho) + e^{h(0)}\rho).$$

Note that the right-hand side is strictly greater than $\gamma + \log(e^{h(1)}(1 - \rho))$, which equals to the left-hand side. Similar calculations for case (II) also lead to a contradiction. Hence, although an optimal cost is constant, the optimality inequality need not have a solution.

Now we turn to checking Condition (**B**). Let $V_\beta$ be as in Lemma 2. Clearly, $V_\beta = w_\beta^N$ for $N \geq 1$ and $V_\beta(0) = 0$. Then, by (8) under (I), we get

$$V_\beta(1) = \gamma + \log[e^{\beta V_\beta(1)}(1 - \rho) + \rho] < \gamma + \log[e^{V_\beta(1)}(1 - \rho) + \rho].$$

Hence,
$$V_\beta(1) < \log\left(\frac{e^\gamma(1-\rho)}{1-e^\gamma(1-\rho)}\right) \qquad \forall \beta \in (0,1),$$
and consequently, $\sup_{\beta \in (0,1)} h_\beta(x) < +\infty$.

Now let the risk factor $\gamma$ be as in (III). Then by (8),
$$V_\beta(1) > \gamma + \log(1-\rho) + \beta V_\beta(1),$$
which in turn implies that
$$V_\beta(1) > \frac{\gamma + \log(1-\rho)}{1-\beta}.$$
Thus, $h_\beta(1) = V_\beta(1)$ goes to the infinity when $\beta \nearrow 1$.

For case (II), we obtain
$$\begin{aligned}
V_\beta(1) &= -\log(1-\rho) + \log[e^{\beta V_\beta(1)}(1-\rho) + \rho] \\
&= \beta V_\beta(1) + \log\left[1 + e^{-\beta V_\beta(1)}\frac{\rho}{1-\rho}\right].
\end{aligned} \qquad (31)$$

If $V_\beta(1) \nearrow +\infty$ when $\beta \nearrow 1$, then the right-hand side of (31) also goes to the infinity. On the contrary, assume that $\sup_{\beta \in (0,1)} V_\beta(1) \leq C$ for some constant $C > 0$. Then,
$$V_\beta(1) \geq \frac{\log[1 + e^{-C}\rho/(1-\rho)]}{1-\beta},$$
which leads to a contradiction when $\beta \nearrow 1$. In consequence, in case (II) the family $\{h_\beta(1)\}$ does not satisfy Condition (B) either.

Therefore, the following conclusion can be drawn. Condition (B) is necessary to obtain a solution to the optimality inequality.

For a verification of Condition (B), one can use Lemma 4 below. For a similar result in the risk-neutral, case we refer to [27, 28]. For some $\eta \geq 0$, define the stopping time
$$\tau = \tau(\beta) := \inf\{n \geq 0 : V_\beta(x_n) \leq m_\beta + \eta\}.$$

LEMMA 4. *For $\eta \geq 0$, $\beta \in (0,1)$ and $x \in X$,*
$$h_\beta(x) \leq \eta + \inf_{\pi \in \Pi} \log E_x^\pi \exp\left(\sum_{k=0}^{\tau-1} \gamma c(x_k, a_k)\right).$$

PROOF. By Lemma 2(b), (c) and the fact that $V_\beta(y) \geq 0$, $y \in X$, we have
$$\begin{aligned}
V_\beta(x) &= \min_{a \in A(x)}\left[\gamma c(x,a) + \log \int_X e^{\beta V_\beta(y)} q(dy|x,a)\right] \\
&< \gamma c(x,a) + \log \int_X e^{V_\beta(y)} q(dy|x,a)
\end{aligned} \qquad (32)$$

for each $x \in X$. Subtracting $m_\beta$ from both sides in (32), we obtain

$$V_\beta(x) - m_\beta < \gamma c(x,a) + \log \int_X e^{(V_\beta(y)-m_\beta)} q(dy|x,a).$$

Iteration of this inequality up to the stopping time $\tau$ yields

$$V_\beta(x) - m_\beta < \log E_x^\pi e^{\gamma \sum_{k=0}^{\tau-1} c(x_k,a_k) + \eta}$$

$$= \eta + \log E_x^\pi \exp\left(\gamma \sum_{k=0}^{\tau-1} c(x_k, a_k)\right).$$

Since $\pi \in \Pi$ is an arbitrary policy, we easily get the conclusion. □

Note that the fact

(33) $$E_x^\pi \exp\left(\sum_{k=0}^{\tau-1} \gamma c(x_k, a_k)\right) < +\infty$$

has the following interpretation: before the process will reach "good states," the incurred costs at "early stages" should not be too large. Indeed, let us define a set $D$ as follows. We say that

$$x \in D \quad \text{iff} \quad V_\beta(x) \leq m_\beta + \eta$$

for a certain $\eta \geq 0$. Clearly, $D \neq \varnothing$. Denote by $\tau_D$ the first return time of the process, governed by $f_\beta$, to set $D$. Certainly, if (33) holds with $\tau := \tau_D$, then Condition (**B**) is satisfied.

In Example 1 we can take $D = \{0\}$ and $\eta = 0$, since $V_\beta(0) \leq 0 + 0$. If $\gamma$ is as in (I), then (33) holds:

$$E_1 \exp\left(\sum_{k=0}^{\tau_0-1} \gamma c(x_k)\right) = \sum_{n=1}^\infty e^{n\gamma}(1-\rho)^{n-1}\rho = \frac{\rho}{1-\rho}\left(\frac{e^\gamma(1-\rho)}{1-e^\gamma(1-\rho)}\right).$$

In other cases (33) fails to hold and, in addition, the earlier calculations show that $h_\beta(1) = +\infty$.

Summing up, the presented example shows that, without Condition (**B**) imposed on the family of functions $\{h_\beta(x)\}$, $\beta \in (0,1)$, a solution to the optimality inequality need not exist, and moreover, the optimal risk-sensitive average cost may depend on the initial state. In view of the above discussion, Condition (**B**) is designed to prevent the accrual of infinite expected costs. Namely, the costs incurred at transient states, that may be occupied only at "early stages," have an important and definite influence on a long-run performance measure. Therefore, Condition (**B**) requires the model to be sort of communicating insofar as certain sets of "good states" to be reached sufficiently fast. Then, the optimal risk-sensitive average cost is constant and the optimality inequality takes place. In addition, it is worth mentioning that the ergodicity itself of a Markov process/chain does not help so much as in the risk-neutral case. In other words, for an ergodic Markov chain, it may happen that the optimal risk-sensitive average cost depends on the initial state as in Example 1. Moreover, in this example one can even prove in a straightforward way that under case (I) [either under Condition (B) or for sufficiently small risk factors], the optimality equation (24) is satisfied. Therefore, it would be interesting to know whether Condition (B) (together with some compactness–continuity assumptions) is sufficient to obtain a solution to the optimality equation. There is a conjecture that, since in the risk-neutral case a counterpart of Condition (B) is not sufficient [7], neither is it in the risk-sensitive setting. But this question is beyond the scope of the paper and remains open.

## APPENDIX

The lemma below establishes a variational formula for the logarithmic moment-generating function. The reader is referred to Theorem 4.5.1 and Proposition 1.4.2 in [12] for its proof.

LEMMA A. *Let $\mathcal{X}$ be a Polish space, $h$ a measurable function mapping on $\mathcal{X}$ into $\mathbb{R}$, which is either bounded from below or bounded from above, and $\nu$ a probability measure on $\mathcal{X}$.*

(a) *Then, we have the variational formula*

$$\log \int_{\mathcal{X}} e^h d\nu = \sup_{\mu \in \Delta} \left( -R(\mu\|\nu) + \int_{\mathcal{X}} h \, d\mu \right),$$

*where*

$$\Delta = \{\mu \in \Pr(\mathcal{X}) : R(\mu\|\nu) < +\infty\}.$$

(b) *Let $\mu_0$ denote the probability measure on $\mathcal{X}$, which is $\mu_0 \ll \nu$ and satisfies*

$$\frac{d\mu_0}{d\nu}(x) = \frac{e^{h(x)}}{\int_{\mathcal{X}} e^h \, d\nu}.$$

*Then, the supremum in the variational formula is attained uniquely at $\mu_0$.*

**Acknowledgments.** A part of this research was done while the author was a Humboldt research fellow and visiting the University of Ulm. The author gratefully acknowledges support from the Alexander von Humboldt Foundation.

The second part of this paper was written at the Institute of Mathematics and Computer Science, Wrocław University of Technology.

The author is greatly indebted to Professor Ulrich Rieder for drawing her attention to paper [16], suggesting the problem and for several helpful conversations.

Institute of Mathematics and Computer Science
Wrocław University of Technology
Wybrzeże Wyspiańskiego 27
PL-50-370 Wrocław
Poland
E-mail: ajaskiew@im.pwr.wroc.pl